\documentclass[12pt]{article}

\usepackage[english,russian]{babel}
\usepackage{amsfonts}
\usepackage{amscd}
\usepackage{amsmath}
\usepackage{amssymb}
\usepackage{latexsym}
\usepackage{longtable}

\begin{document}

\begin{center}
\large  Non-existence of a short algorithm for multiplication \\ 
 of $3\times3$ matrices with group $S_4\times S_3$ \\
\medskip
\normalsize Vladimir~P.~Burichenko  \\
\medskip
\small Institute of Mathematics of the National Academy of Sciences of Belarus \\
 e-mail: vpburich@gmail.com 
\end{center}
\bigskip

\begin{abstract}\noindent
One of prospective ways to find new fast algorithms of matrix
multiplication is to study algorithms admitting nontrivial
symmetries. In the work possible algorithms for multiplication of
$3\times3$ matrices, admitting a certain group $G$ isomorphic to
$S_4\times S_3$, are investigated. It is shown that there exist no
such algorithms of length $\leq23$. In the first part of the work,
which is the content of the present article, we describe all
orbits of length $\leq23$ of $G$ on the set of decomposable
tensors in the space $M\otimes M\otimes M$, where $M=M_3({\mathbb
C})$ is the space of complex $3\times3$ matrices. In the second
part of the work this description will be used to prove that a
short algorithm with the above-mentioned group does not exist.

({\em MSC classification 68Q25, 20C}).
\end{abstract}

{\bf 1. Introduction.} The fast matrix multiplication is one of
the main questions about computational complexity, see e.g.~[1,2].
It was proposed in works [3,4] and independently in [5] to
consider algorithms admitting nontrivial symmetries. This may be a
prospective way to find new fast algorithms. The exact definition
of what is the automorphism group of an algorithm can be found in
~[4](also in [5], [6]).

In [3,4] the automorphism groups of well-known algorithms of
Strassen, Hopcroft and Laderman were found. These groups are
isomorphic to $S_3\times S_3$, $S_3\times Z_2$, and $S_4$,
respectively. This suggests the following idea: take a group,
which is more or less similar to these three, and investigate
algorithms for multiplication of $3\times3$ matrices that are
invariant under this group (with an intention to find an algorithm
shorter than Laderman's).

Recently several works [7--9, 9a] were published where algorithms
invariant under a prescribed groups were studied. However, no
algorithms better than the known ones (say, an algorithm for
$3\times3$ matrices of length $\leq22$; remind that Laderman's
algorithm is of length $23$. On the other hand, a lower estimate
$\geq19$ for the length of such an algorithm is known, see [10])
were found.

In the present article we consider certain group isomorphic to
$S_4\times S_3$ as a candidate for the symmetry group.
Unfortunately, the result is negative again: an algorithm of
length $\leq23$ with this group does not exist. The details of
computations are nontrivial and may be useful for future research.

For the convenience of the reader who is not especially
experienced in algorithms we formulate our result in purely
representation-theoretic terms. Let
$$ M=M_3({\mathbb C})=\langle e_{ij}\mid 1\leq i,j\leq3\rangle_{\mathbb C}$$
be the space of complex $3\times3$ matrices. Here $e_{ij}$ are the
usual matrix units. Consider the tensor
$$ {\mathcal T}=\sum_{1\leq i,j,k\leq3}e_{ij}\otimes e_{jk}\otimes e_{ki}
\in M\otimes M\otimes M\, $$ (in complexity theory this tensor is
usually denoted by $\langle3,3,3 \rangle$).

Let $A\leq GL(3,{\mathbb C})$ be the group of all monomial
$3\times3$ matrices whose nonzero elements are $\pm1$ and the
determinant is $\det=1$. It is easy to see that $A\cong S_4$ and
$A$ is irreducible.

For any $a,b,c\in GL(3,{\mathbb C})$ consider the transformation
of $M\otimes M\otimes M$,
$$ T(a,b,c)\colon x\otimes y\otimes z\mapsto axb^{-1}\otimes byc^{-1}
\otimes cza^{-1}\,.$$ It is easy to see that always
$T(a,b,c){\mathcal T}={\mathcal T}$. In particular,
$T(a,a,a){\mathcal T}={\mathcal T}$ for $a\in A$. Thus we can
consider that $A$ acts on $M^{\otimes3}$ and preserves ${\mathcal
T}$.

Also, consider the two transformations
$$ \rho(x\otimes y\otimes z)=y^t\otimes x^t\otimes z^t\,,\qquad
\sigma(x\otimes y\otimes z)=z\otimes x\otimes y\,$$ (where $t$
means transpose). It is easy to see that both $\rho$ and $\sigma$
preserve ${\mathcal T}$, and that $B:=\langle\rho,\sigma\rangle
\cong S_3$. Finally, it is not hard to see that $A$ and $B$
commute elementwise (for the details of these (and more general)
computations the reader is referred to [4] or~[6]). Thus, the
group $G=A\times B\cong S_4\times S_3$ acts on $M^{\otimes3}$ and
preserves ${\mathcal T}$.

A {\em decomposition of length $l$} of ${\mathcal T}$ is an
(unordered) set of $l$ decomposable tensors $\{x_i\otimes
y_i\otimes z_i\mid i=1, \ldots,l\}$ such that
$$ \sum_{i=1}^l x_i\otimes y_i\otimes z_i={\mathcal T}\,.$$
(Note that in the previous sentence the word ``decomposition'' was
used twice, first in ``additive'' sense, then in the
multiplicative one !).

Clearly, any element of $G$ takes a length $l$ decomposition into
a length $l$ decomposition also. So we can consider {\em
$G$-invariant} decompositions. Now we can state the main result of
the work.
\smallskip

{\bf Theorem 1.} {\em Let ${\mathcal T}=\langle3,3,3\rangle$ and
$G=A\times B$ be as described above. Then there exists no
$G$-invariant decomposition of ${\mathcal T}$ of length $\leq23$.
}
\smallskip

We divide the proof of this theorem into two parts. In the first
part, which is the content of the present article, we describe the
orbits of $G$ on decomposable tensors in $M^{\otimes3}$, of length
$\leq23$ (this description turns out to be rather long). In the
second part, which will be published soon, we make use of this
description to prove Theorem~1.

The reader should be warned that this proof contains extensive
calculations. To write all these calculations in full would be
tiresome. So, we usually only explain the main idea of a
calculation and give an example or two, and leave the rest of
calculations to an interested reader.

{\bf Remark.} Observe that the standard decomposition of
${\mathcal T}$, that is, the set of all $e_{ij}\otimes
e_{jk}\otimes e_{ki}$, is $G$-invariant (but its full automorphism
group is much larger, namely of the form $({\mathbb
C}^\ast)^6\leftthreetimes S_3$). It is interesting to find out
whether there are $G$-invariant decompositions of smaller length
(say, $26$).

\smallskip

{\bf 2. The subgroups of $S_4\times S_3$.} Thus, the main aim of
the present article is to classify the orbits of $G$ on the
decomposable tensors $x\otimes y\otimes z$, of length $\leq23$. In
fact, the length of such an orbit is $\leq18$, because $S_4\times
S_3$ has no subgroups of index $19,20,21$, $22$, or $23$. In this
section we describe all subgroups of index $\leq18$ of  $S_4\times
S_3$. We write elements of $S_4\times S_3$ as pairs of
permutations. (The explicit correspondence between permutation
notation and action of $G$ on $M\otimes M\otimes M$ will be
described in the next section). We consider the permutations as
acting on symbols $1,2,3,4$ on the left (and so multiplied from
right to left, like $(12)(13)=(132)$).

The subgroups of $S_3$ are, up to isomorphism, $1$, $Z_2$, $Z_3$,
and $S_3$ itself. And two isomorphic subgroups are always
conjugated. So we denote the conjugacy class of a subgroup by the
same symbol as its isomorphism class. Also, the subgroups of $S_4$
are, up to conjugacy, $1$, $Z_2^{(1)}$, $Z_2^{(2)}$, $Z_3$,
$V^{(1)}$, $V^{(2)}$, $Z_4$, $S_3$, $D_8$, $A_4$, and $S_4$. Here
$V=Z_2\times Z_2$, and the superscript is used to distinguish
between non-cojugate isomorphic subgroups:
$$ Z_2^{(1)}\sim \langle(12)\rangle,\qquad Z_2^{(2)}\sim \langle(12)(34)
\rangle, $$
$$ V^{(1)}\sim \langle(12), (34)\rangle,\qquad V^{(2)}\sim \langle(12)(34),
(13)(24)\rangle\,.$$

Recall that a {\em subdirect product} of groups $X$ and $Y$ is any
subgroup $Z\leq X\times Y$ such that $\pi_X(Z)=X$ and
$\pi_Y(Z)=Y$, where $\pi_X$ and $\pi_Y$ are the projections of
$X\times Y$ onto factors. The subdirect products can be
characterized by the following property (see [11], \S 5.5): there
exists a group $W$, which is a quotient group for both $X$ and
$Y$, and two epimorphism  $\varphi_X\:\colon\:X\longrightarrow W$
and $\varphi_Y\:\colon\:Y\longrightarrow W$  such that
$$ Z=\{ (x,y)\in X\times Y\mid \varphi_X(x)=\varphi_Y(y)\}\,.$$
Conversely, any subgroup of this form is a subdirect product of
$X$ and $Y$.

A (nontrivial) subdirect product of $X$ and $Y$ will be denoted by
$X\circ Y$ (or $X\circ_i Y$, if there are several such nontrivial
products).

If $X$ and $Y$ are arbitrary groups and $Z$ any subgroup of
$X\times Y$, then clearly $Z$ is a subdirect product of the
projections $X_1=\pi_X(Z)$ and $Y_1=\pi_Y(Z)$. Moreover, if
$X_1,X_2\leq X$ are conjugate in $X$, and $Y_1,Y_2\leq Y$ are
conjugate also, and $Z_1$ is a subdirect product of $X_1$ and
$Y_1$, then $Z_1$ is conjugate in $X\times Y$ with a subgroup
$Z_2$ that is a subdirect product of $X_2$ and $Y_2$.

Note also that if $C$ and $D$ are subgroups of $X$ and $Y$
respectively, determined up to conjugacy, then $C\times D$ can be
considered as a subgroup of $X\times Y$, determined also up to
conjugacy.

For each subgroup of $S_4$ or $S_3$ to find all its quotients is
trivial. Now, keeping in mind the previous discussion, it is easy
to prove the following statement.
\smallskip

{\bf Proposition 2.} {\em All the subgroups of $S_4\times S_3$ of
index $\leq18$ (that is, of order $\geq8$) are the following, up
to conjugacy: \vspace{-1.0ex}
$$Z_2^{(1)}\times S_3\,,\quad Z_2^{(2)}\times S_3\,,\quad Z_3\times Z_3\,,
\quad Z_3\times S_3\,,\quad V^{(1)}\times Z_2\,,\quad
V^{(1)}\times Z_3\,, $$ \vspace{-4.0ex}
$$V^{(1)}\times S_3\,,\quad V^{(1)}\circ_i S_3\ (i=1,2)\,,\quad
V^{(2)}\times Z_2\,,\quad V^{(2)}\times Z_3\,,\quad V^{(2)}\times
S_3\,,$$ \vspace{-3.5ex}
$$ V^{(2)}\circ S_3\,,\quad Z_4\times Z_2\,,\quad Z_4\times Z_3\,,\quad
Z_4\times S_3\,,\quad Z_4\circ S_3\,,\quad S_3\times Z_2\,,\quad
S_3\times Z_3\,,$$ \vspace{-3.0ex}
$$ S_3\times S_3\,,\quad S_3\circ S_3\,,\quad D_8\times1\,,\quad
D_8\times Z_2\,,\quad D_8\times Z_3\,,\quad D_8\times S_3\,,$$
\vspace{-3.0ex}
$$D_8\circ_i Z_2\ (i=1,2,3)\,,\quad
D_8\circ_i S_3\,,\quad A_4\times1\,,\quad  A_4\times Z_2\,,\quad
A_4\times Z_3\,, \quad A_4\times S_3\,,$$ \vspace{-3.0ex}
$$ \quad A_4\circ Z_3\,,\quad S_4\times1\,,\quad S_4\times Z_2\,,\quad
S_4\times Z_3\,,  S_4\times S_3\,, \quad S_4\circ Z_2\,,\quad
S_4\circ_i S_3\ (i=1,2).$$ }

{\bf Sketch of proof.} We know from the previous discussion that
any subgroup $Z\leq S_4\times S_3$ is a subdirect product of
$X\leq S_4$ and $Y\leq S_3$. And we can take $X$ and $Y$ up to
conjugacy, if we are interested in the conjugacy class of $Z$
only.

Consider, for example, the case $X=Z_4$, $Y=S_3$. Notice that the
only nontrivial quotient of both $Z_4$ and $S_3$ is $Z_2$, and the
epimorphisms $Z_4\longrightarrow Z_2$ and $S_3\longrightarrow Z_2$
are unique. So there exists a unique nontrivial subdirect product
of $Z_4$ and $S_3$. Thus, in the case under consideration we
obtain two possibilities for $Z$: $Z=Z_4\times S_3$ or $Z=Z_4\circ
S_3$.

In some cases the subdirect product is not unique. For example,
let $X=D_8$ and  $Y=S_3$. The unique quotient of both $X$ and $Y$
is $Z_2$. The epimorphism $S_3\longrightarrow Z_2$ is unique, but
there are three distinct epimorphisms $D_8\longrightarrow Z_2$. So
there are three distinct subdirect products $D_8\circ_i S_3$,
$i=1,2,3$. A similar argument applies for subgroups of the form
$D_8\circ_i Z_2$.

We mention also the case of groups $S_4\circ S_3$. A common
quotient of $S_4$ and $S_3$ is either $Z_2$ or $S_3$. In the case
$Z_2$ the corresponding subgroup is unique, because the
epimorphisms $S_4\longrightarrow Z_2$ and $S_3\longrightarrow Z_2$
are unique. In the case $S_3$ the subgroup is not unique, since
there are distinct epimorphisms $S_4\longrightarrow S_3$ and
$S_3\longrightarrow S_3$. However, since any automorphism of $S_3$
is an inner one, all these subgroups are conjugate in $S_4\times
S_3$ (by an element of the form $(1,x)$, where $x\in S_3$). The
subdirect product with common quotient $Z_2$ will be denoted by
$S_4\circ_1 S_3$, and that with quotient $S_3$ by $S_4\circ_2
S_3$. Similarly, the subgroups of the form $A_4\circ Z_3$ are
conjugate also (there exist two epimorphisms $A_4\longrightarrow
Z_3$, but the corresponding subdirect products $A_4\circ_1Z_3$ and
$A_4\circ_2Z_3$ are conjugate by an element of the form $(1,y)$,
where $y\in S_3$ is a transposition).

Finally, consider subdirect products of the form $V^{(j)}\circ
S_3$, $j=1,2$. Since there are three distinct epimorphisms
$V^{(j)}\longrightarrow Z_2$, there exists three  products in each
of two cases $j=1,2$. However, all three order $2$ subgroups of
$V^{(2)}$ are conjugate under normalizer $N_{S_4}(V^{(2)})$
($=S_4$), and in the case $V^{(1)}$ there are two conjugacy
classes of such subgroups. Hence three products of the form
$V^{(2)}\circ S_3$ are conjugate in $S_4\times S_3$, and there are
two non-conjugate products of the form $V^{(1)}\circ S_3$. \hfill
$\square$
\medskip

Now, we choose some representatives of conjugacy classes of
subgroups of $S_3$ and $S_4$, and, next, of subgroups of
$S_4\times S_3$.

As representatives for $Z_2$ and $Z_3$ in $S_3$ we take
$\langle(12)\rangle_2$ and $\langle(123)\rangle_3$ respectively.

When choosing representatives for conjugacy classes of subgroups
of $S_4$ we take care to take them in such a way that there will
be inclusions among them, in natural situations. The
representatives for $Z_2^{(1)}$, $Z_2^{(2)}$, $V^{(1)}$, and
$V^{(2)}$ are chosen already. Then it is natural to take $\langle
V^{(2)}, Z_2^{(1)}\rangle=\langle (12)(34), (13)(24), (12)\rangle$
as a representative of $D_8$; note that it contains the previous
four groups. Now, it is natural to take the only subgroup
isomorphic to $Z_4$ in the latter group, namely
$\langle(1324)\rangle_4$, as a representative for $Z_4$. Next, as
a representative of $Z_3$ take $\langle(123)\rangle$. Then
$S_3=\langle Z_3, Z_2^{(1)}\rangle$ is the subgroup of all
permutations preserving $\{1,2,3\}$ and fixing $4$. Finally, there
is no need to choose representatives for $A_4$ and $S_4$, since
such subgroups are unique. (In fact, there was no need to choose a
representative for a class of subgroups conjugate to $V^{(2)}$
also, because $V^{(2)}$ is normal in $S_4$.)

In the sequel, when mentioning a subgroup $X\times Y\leq S_4\times
S_3$, we mean that $X$ and $Y$ are the standard representatives of
their conjugacy classes, and similarly for $X\circ Y$.

Now we can describe more precisely the structure of nontrivial
subdirect products in Proposition~2. Let $H=X\circ Y$ be such a
product and $C=H\cap A$, $D=H\cap B$ (or, more exactly, $C$ is a
subgroup of $A$, corresponding (with respect to the isomorphism
$A\cong A\times1$) to the subgroup $H\cap(A\times1)$; and
similarly for $D$). Then $H\geq C\times D$, and $X/C\cong Y/D\cong
H/C\times D$ is a common quotient for $X$ and $Y$. If there exist
$R_1\leq X$ and $R_2\leq Y$ such that $X=C\leftthreetimes R_1$ and
$Y=D\leftthreetimes R_2$ (then necessarily $R_1\cong R_2\cong
X/C$), then $H=(C\times D)\leftthreetimes R$, where $R$ is a
diagonal subgroup of $R_1\times R_2$.

In fact, the only case of all the cases listed in Proposition~2
when $R_1$ or $R_2$ do not exist is the group $Z_4\circ S_3$.
Also, in most cases $R\cong Z_2$. More exactly, the reader can
easy verify the following proposition.
\smallskip

{\bf Proposition 3.} {\em The following equalities hold:
$$ V^{(1)}\circ_1 S_3=(Z_2^{(1)}\times Z_3)\leftthreetimes \langle g_1
  \rangle_2\,, \qquad  V^{(1)}\circ_2 S_3=(Z_2^{(2)}\times Z_3)
  \leftthreetimes \langle g_2\rangle_2\,, $$
$$ V^{(2)}\circ S_3=(Z_2^{(2)}\times Z_3)\leftthreetimes \langle
  g_3\rangle_2\,, \qquad Z_4\circ S_3=(Z_2^{(2)}\times Z_3) \langle g_4
  \rangle_4\,, $$
(note nontrivial intersection in the latter case !),
$$ S_3\circ S_3=(Z_3\times Z_3)\leftthreetimes \langle g_2\rangle_2\,,
\qquad  D_8\circ_1 Z_2=(V^{(1)}\times 1)\leftthreetimes \langle
g_3\rangle_2\,, $$
$$ D_8\circ_2 Z_2=(V^{(2)}\times 1)\leftthreetimes \langle g_2\rangle_2\,,
   \qquad  D_8\circ_3 Z_2=(Z_4\times 1)\leftthreetimes
	\langle g_2\rangle_2\,, $$
$$ D_8\circ_1 S_3=(V^{(1)}\times Z_3)\leftthreetimes \langle g_3\rangle_2\,,
	\qquad  D_8\circ_2 S_3=(V^{(2)}\times Z_3)\leftthreetimes
	\langle g_2\rangle_2\,, $$
$$ D_8\circ_3 S_3=(Z_4\times Z_3)\leftthreetimes \langle g_2\rangle_2\,,
   \qquad  A_4\circ Z_3=(V^{(2)}\times 1)\leftthreetimes
   \langle g_5\rangle_3\,, $$
$$ S_4\circ Z_2=(A_4\times 1)\leftthreetimes \langle g_2\rangle_2\,, \qquad
   S_4\circ_1 S_3=(A_4\times Z_3)\leftthreetimes \langle g_2\rangle_2\,, $$
$$ S_4\circ_2 S_3=(V^{(2)}\times 1)\leftthreetimes
   \langle g_2,g_5\rangle_2\,, $$
where
$$ g_1=((12)(34), (12)), \quad g_2=((12), (12)), \quad g_3=((13)(24),(12)), $$
$$ g_4=((1324), (12)), \quad g_5=((123),(123))\,.$$
\hfill $\square$ }
\smallskip

(Note that in this proposition we implicitly introduced (and shall
use in the sequel) a numbering for the products $X\circ_i Y$ in
the cases where there exist several products of the form $X\circ
Y$).
\smallskip

{\bf 3. Correspondences $A\leftrightarrow S_4$, $B\leftrightarrow
S_3$.} For further considerations we need to choose isomorphisms
$A\leftrightarrow S_4$ and $B\leftrightarrow S_3$ in an explicit
form.

The isomorphism $B\leftrightarrow S_3$ is taken in such a way that
the tensor factors of $M\otimes M\otimes M$ are permuted in the
same way as the symbols $1,2,3$, that is
$$ e\leftrightarrow (x\otimes y\otimes z\mapsto x\otimes y\otimes z),\qquad
   (12)\leftrightarrow (x\otimes y\otimes z\mapsto y^t\otimes x^t\otimes z^t),$$
$$ (13)\leftrightarrow (x\otimes y\otimes z\mapsto z^t\otimes y^t\otimes x^t),
   \qquad  (23)\leftrightarrow (x\otimes y\otimes z\mapsto x^t\otimes z^t
   \otimes y^t), $$
$$ (123)\leftrightarrow (x\otimes y\otimes z\mapsto z\otimes x\otimes y),
	\qquad  (132)\leftrightarrow (x\otimes y\otimes z\mapsto y\otimes z
	 \otimes x). $$
This formulae mean the following: the element $(12)\in S_3$
corresponds to the transformation $x\otimes y\otimes z\mapsto
y^t\otimes x^t\otimes z^t$ of $B$, etc. The fact that this is an
isomorphism indeed easily follows from the observation that the
permutation of factors in the tensor product commutes with taking
componentwise transpose, i.e. with transformation $x\otimes
y\otimes z\mapsto x^t\otimes y^t\otimes z^t$.

Next we describe an isomorphism between $A$ and $S_4$. First let
$g\in S_3\leq S_4$, $\widetilde g\in GL(3,{\mathbb C})$ be the
corresponding permutation matrix (i.e. $\widetilde ge_i=e_{gi}$,
for all $i=1,2,3$, whence $\widetilde
g=e_{g1,1}+e_{g2,2}+e_{g3,3}$; here $e_1$, $e_2$, $e_3$ are unit
column vectors (one of whose entries is $1$, the others $0$)), and
$\widehat g={\rm sgn}(g)\cdot\widetilde g=\pm\widetilde g$. It is
clear that $\widehat g\in A$, and the correspondence
$\alpha\:\colon\: g\mapsto \widehat g$ is an injective
homomorphism from $S_3$ to~$A$. Its image will be denoted by
$\widehat S_3$.

It is not hard to extend $\alpha$ to an isomorphism of $S_4$
onto~$A$. Observe that
$$ C=\{{\rm diag}(\varepsilon_1,\varepsilon_2,\varepsilon_3)\mid
\varepsilon_{1,2,3}=\pm1,\
\varepsilon_1\varepsilon_2\varepsilon_3=1\} $$ is a normal
subgroup of $A$ isomorphic to $Z_2^2$ and $A=C\leftthreetimes
\widehat S_3$. On the other hand, $V^{(2)}$ is normal in $S_4$ and
isomorphic to $Z_2^2$ also, and $S_4=V^{(2)}\leftthreetimes S_3$.
The unique element of $V^{(2)}$ commuting with $(12)$ is
$(12)(34)$; the unique element of $C$ commuting with
$\widehat{(12)}=-e_{12}-e_{21}-e_{33}$ is ${\rm diag}(-1,-1,1)$.
This suggests an idea to consider the correspondence
$\beta\:\colon\:V^{(2)}\longrightarrow C$ defined by
$$ \beta\:\colon\: (12)(34)\mapsto{\rm diag}(-1,-1,1)=\widehat{(12)}, \qquad
  (13)(24)\mapsto\widehat{(13)}={\rm diag}(-1,1,-1), $$
$$  (14)(23)\mapsto\widehat{(23)}={\rm diag}(1,-1,-1), \qquad
	e\mapsto E={\rm diag}(1,1,1). $$

Now it is possible to show that the map
$\gamma\:\colon\:S_4\longrightarrow A$ defined by
$\gamma(xy)=\beta(x)\alpha(y)$, where $x\in V^{(2)}$ and $y\in
S_3$, is an isomorphism (the details are left to the reader. For
example, we may check that $\gamma$ preserves defining relations
for a certain system of generators of $S_4$).

We conclude this section with a very useful remark.

{\bf Remark.} {\em Let $\pi\in S_3\leq S_4$. Then action of an
element $\widehat\pi=\alpha(\pi)\in A$ on $M$ coincides with ``the
action of $\pi$ on indices'', that is $\pi e_{ij}=\widehat\pi
e_{ij}\widehat\pi^{-1} =e_{\pi i,\pi j}$. }
\smallskip

{\bf 4. The theorem on classification of $G$-orbits.} Now we can
state the main result of the present article.

Below in the article ${\rm St}_G(w)$ denotes the stabilizer of a
tensor $w$ under the action of $G$; ``subvariety'' means an
algebraic subset (i.e., closed in Zariski topology); $\zeta$ is a
primitive cubic root of $1$, $i=\sqrt{-1}$ (also, we use the same
letter $i$ for indices, but hope this will not cause a confusion
even in the formulae like
 $e_{ij}-ie_{ki}$). Next,
$$ \delta=e_{11}+e_{22}+e_{33}\,,\quad \varkappa=\sum_{i\ne j}e_{ij}=
   e_{12}+e_{21}+e_{13}+e_{31}+e_{23}+e_{32}\,,$$
$$\eta=e_{11}+\zeta e_{22}+\overline\zeta e_{33}\,,\qquad \overline\eta=
   e_{11}+\overline\zeta e_{22}+   \zeta e_{33}\,,$$
$$\tau=e_{12}+e_{23}+e_{31}-e_{21}-e_{32}-e_{13}\,. $$

{\bf Theorem 4.} {\em The orbits of the group $G=A\times B$ on the
set of nonzero decomposable tensors in $M^{\otimes3}$ of length
$\leq18$ are described by the data written in the table below, in
the following sence. If ${\mathcal O}$ is an orbit of length
$\leq18$, then there exists a point (i.e., a tensor)
$w\in{\mathcal O}$, a row $(i,H_i,l_i,w_i(a,b,\ldots))$ of a
table, and (in general, not uniquely determined) parameters
$a,b,\ldots\in{\mathbb C}$ such that $w=w_i(a,b,\ldots)$, ${\rm
St}_G(w)=H_i$, and $|{\mathcal O}|=l_i$. Conversely, for each row
$(i,H_i,l_i,w_i(a,b,\ldots))$ there exists a proper subvariety
$Q_i\subset{\mathbb C}^{s_i}$ (where $s_i$ is the number of
parameters $a,b,\ldots$, for given $i$) such that for
$x=(a,b,\ldots)\in{\mathbb C}^{s_i}\setminus Q_i$ the orbit
${\mathcal O}$ of $w=w_i(x)$ has length $l_i$ and ${\rm
St}_G(w)=H_i$. If $x\in Q_i$, then ${\rm St}_G(w)$ strictly
contains $H_i$, and the length of its orbit is $<l_i$.

\begin{longtable}{|p{0.5cm}|p{2cm}|p{0.5cm}|p{12cm}|}
\hline
\rule{0pt}{3ex}  $i$ & $H_i$ & $l_i$ & $w_i(a,b,\ldots)$ \\
\hline  \endhead
  $1$ & $Z_2^{(1)}\times S_3$ & $12$ & \rule[-2ex]{0pt}{5ex}
   $(a(e_{11}+e_{22})+b(e_{12}+e_{21})+ce_{33}+d(e_{13}+e_{23}+
   e_{31}+e_{32}))^{\otimes3}$ \\
\hline  $2$ & $Z_2^{(2)}\times S_3$ & $12$ & \rule[-2ex]{0pt}{5ex}
  $ (ae_{11}+be_{22}+ce_{33}+d(e_{12}+e_{21}))^{\otimes3}$ \\
\hline $3$   & $V^{(1)}\times S_3$ & $6$ & \rule[-2ex]{0pt}{5ex}
   $ (a(e_{11}+e_{22})+be_{33}+c(e_{12}+e_{21}))^{\otimes3}$ \\
\hline $4$  & $V^{(2)}\times S_3$ & $6$ & \rule[-2ex]{0pt}{5ex}
   $(ae_{11}+be_{22}+ce_{33})^{\otimes3}$ \\
\hline $5$ & $D_8\times S_3$ & $3$ & \rule[-2ex]{0pt}{5ex}
	$(a(e_{11}+e_{22})+be_{33})^{\otimes3}$ \\
\hline $6$ & $A_4\times S_3$ & $2$ & \rule[-2ex]{0pt}{5ex}
   $a\eta^{\otimes3}$ \\
\hline $7$ & $S_4\times S_3$ & $1$ & \rule[-2ex]{0pt}{5ex}
	   $a\delta^{\otimes3}$ \\
\hline $8$  & $Z_3\times Z_3$ & $16$ & \rule[-2ex]{0pt}{5ex}
	$(a\eta+b(e_{12}+\zeta e_{23}+\overline\zeta e_{31})
	+c(e_{21}+\zeta e_{32}+\overline\zeta e_{13}))^{\otimes3}$ \\
\hline $9$ & $S_3\times S_3$ & $4$ & \rule[-2ex]{0pt}{5ex}
	   $(a\delta+b\varkappa)^{\otimes3}$ \\
\hline $10$  & $Z_3\times S_3$ & $8$ & \rule[-2ex]{0pt}{5ex}
	$(a\eta+b(e_{12}+e_{21}+\zeta(e_{23}+e_{32})+
	\overline\zeta(e_{31}+e_{13})))^{\otimes3}$ \\
\hline $11$  & $S_3\circ S_3$ & $8$ & \rule[-2ex]{0pt}{5ex}
	$(a\delta+b(e_{12}+e_{23}+e_{31})+c(e_{21}+e_{32}+e_{13}))^{\otimes3}$ \\
\hline $12$ & $D_8\circ_3 S_3$ & $6$ & \rule[-2ex]{0pt}{5ex}
   $(a(e_{11}+e_{22})+b(e_{12}-e_{21})+ce_{33})^{\otimes3}$ \\
\hline $13$ & $V^{(1)}\circ_1 S_3$ & $12$ & \rule[-2ex]{0pt}{5ex}
   $(a(e_{11}+e_{22})+b(e_{12}+e_{21})+ce_{33}+d(e_{13}+e_{23}
	-e_{31}-e_{32}))^{\otimes3}$ \\
\hline $14$ & $V^{(1)}\circ_2 S_3$ & $12$ & \rule[-2ex]{0pt}{5ex}
   $(a(e_{11}+e_{22})+be_{12}+ce_{21}+de_{33})^{\otimes3}$ \\
\hline $15$ & $V^{(2)}\circ S_3$ & $12$ & \rule[-2ex]{0pt}{5ex}
   $(ae_{11}+be_{22}+c(e_{12}-e_{21})+de_{33})^{\otimes3}$ \\
\hline  $16$ & $D_8\times1$ & $18$ & \rule[-2ex]{0pt}{5ex}
   $(a(e_{11}+e_{22})+be_{33})\otimes (c(e_{11}+e_{22})+de_{33})\otimes
	(f(e_{11}+e_{22})+ge_{33})$     \\
\hline $17$ & $D_8\times1$ & $18$ & \rule[-2ex]{0pt}{5ex}
$a(e_{11}-e_{22})
	 \otimes(e_{12}+e_{21})\otimes(e_{12}-e_{21})$  \\
\hline $18$ & $D_8\times Z_2$ & $9$ & \rule[-2ex]{0pt}{5ex}
	   $(e_{11}-e_{22})^{\otimes2}\otimes(a(e_{11}+e_{22})+be_{33})$  \\
\hline $19$ & $D_8\times Z_2$ & $9$ & \rule[-2ex]{0pt}{5ex}
	   $(e_{12}+e_{21})^{\otimes2}\otimes(a(e_{11}+e_{22})+be_{33})$   \\
\hline $20$  & $D_8\times Z_2$ & $9$ & \rule[-2ex]{0pt}{5ex}
	  $(e_{12}-e_{21})^{\otimes2}\otimes(a(e_{11}+e_{22})+be_{33})$  \\
\hline $21$  & $D_8\times Z_2$ & $9$ & \rule[-2ex]{0pt}{5ex}
  $(a(e_{11}+e_{22})+be_{33})^{\otimes2}\otimes(c(e_{11}+e_{22})+de_{33})$ \\
\hline $22$   & $V^{(1)}\times Z_2$ & $18$ & \rule[-2ex]{0pt}{5ex}
   $(a(e_{11}+e_{22})+b(e_{12}+e_{21})+ce_{33})^{\otimes2}\otimes
   (d(e_{11}+e_{22})+f(e_{12}+e_{21})+ge_{33})$   \\
\hline $23$   & $V^{(1)}\times Z_2$ & $18$ & \rule[-2ex]{0pt}{5ex}
	$(a(e_{11}-e_{22})+b(e_{12}-e_{21}))\otimes (a(e_{11}-e_{22})-b(e_{12}
	-e_{21}))\otimes (c(e_{11}+e_{22})+d(e_{12}+e_{21})+fe_{33})$  \\
\hline $24$   & $V^{(1)}\times Z_2$ & $18$ & \rule[-2ex]{0pt}{5ex}
	$(a(e_{13}+e_{23})+b(e_{31}+e_{32}))\otimes (b(e_{13}+e_{23})+a(e_{31}
	+e_{32}))\otimes (c(e_{11}+e_{22})+d(e_{12}+e_{21})+fe_{33})$  \\
\hline $25$   & $V^{(2)}\times Z_2$ & $18$ & \rule[-2ex]{0pt}{5ex}
   $(ae_{11}+be_{22}+ce_{33})^{\otimes2}\otimes (de_{11}+fe_{22}+ge_{33})$ \\
\hline $26$   & $V^{(2)}\times Z_2$ & $18$ & \rule[-2ex]{0pt}{5ex}
   $(ae_{12}+be_{21})\otimes(be_{12}+ae_{21})\otimes
   (ce_{11}+de_{22}+fe_{33})$ \\
\hline $27$   & $Z_4\times Z_2$ & $18$ & \rule[-2ex]{0pt}{5ex}
   $(a(e_{11}+e_{22})+b(e_{12}-e_{21})+ce_{33})\otimes (a(e_{11}+e_{22})
   -b(e_{12}-e_{21})+ce_{33})\otimes (d(e_{11}+e_{22})+fe_{33})$ \\
\hline $28$   & $Z_4\times Z_2$ & $18$ & \rule[-2ex]{0pt}{5ex}
   $(a(e_{11}-e_{22})+b(e_{12}+e_{21}))^{\otimes2}\otimes (c(e_{11}+e_{22})
   +de_{33})$ \\
\hline $29$   & $Z_4\times Z_2$ & $18$ & \rule[-2ex]{0pt}{5ex}
   $(a(e_{13}+ie_{23})+b(e_{31}+ie_{32}))\otimes (b(e_{13}+ie_{23})+
	a(e_{31}+ie_{32}))\otimes (c(e_{11}-e_{22})+d(e_{12}+e_{21}))$ \\
\hline $30$   & $D_8\circ_1 Z_2$ & $18$ & \rule[-2ex]{0pt}{5ex}
	$(a(e_{11}+e_{22})+b(e_{12}+e_{21})+ce_{33})\otimes(a(e_{11}+e_{22})
	-b(e_{12}+e_{21})+ce_{33})\otimes (d(e_{11}+e_{22})+fe_{33})$ \\
\hline $31$   & $D_8\circ_1 Z_2$ & $18$ & \rule[-2ex]{0pt}{5ex}
	 $(a(e_{11}-e_{22})+b(e_{12}-e_{21}))^{\otimes2}\otimes(c(e_{11}+
	 e_{22})+de_{33})$ \\
\hline $32$   & $D_8\circ_1 Z_2$ & $18$ & \rule[-2ex]{0pt}{5ex}
	 $(a(e_{13}+e_{23})+b(e_{31}+e_{32}))\otimes (b(e_{13}-e_{23})+
	 a(e_{31}-e_{32}))\otimes (c(e_{11}-e_{22})+d(e_{12}-e_{21}))$ \\
\hline $33$   & $D_8\circ_2 Z_2$ & $18$ & \rule[-2ex]{0pt}{5ex}
	 $(ae_{11}+be_{22}+ce_{33})\otimes (be_{11}+ae_{22}+ce_{33})\otimes
	 (d(e_{11}+e_{22})+fe_{33})$ \\
\hline $34$   & $D_8\circ_2 Z_2$ & $18$ & \rule[-2ex]{0pt}{5ex}
	 $(ae_{12}+be_{21})^{\otimes2}\otimes(c(e_{11}+e_{22})+de_{33})$ \\
\hline $35$   & $D_8\circ_2 Z_2$ & $18$ & \rule[-2ex]{0pt}{5ex}
	 $(ae_{13}+be_{31})\otimes(be_{23}+ae_{32})\otimes(ce_{12}+de_{21})$ \\
\hline $36$   & $D_8\circ_3 Z_2$ & $18$ & \rule[-2ex]{0pt}{5ex}
	 $(a(e_{11}+e_{22})+b(e_{12}-e_{21})+ce_{33})^{\otimes2}\otimes
	 (d(e_{11}+e_{22})+f(e_{12}-e_{21})+ge_{33})$ \\
\hline $37$   & $D_8\circ_3 Z_2$ & $18$ & \rule[-2ex]{0pt}{5ex}
	 $(a(e_{11}-e_{22})+b(e_{12}+e_{21}))\otimes(a(e_{11}-e_{22})-b(e_{12}
	 +e_{21}))\otimes (c(e_{11}+e_{22})+d(e_{12}-e_{21})+fe_{33})$ \\
\hline $38$   & $D_8\circ_3 Z_2$ & $18$ & \rule[-2ex]{0pt}{5ex}
	  $(a(e_{13}+ie_{23})+b(e_{31}+ie_{32}))\otimes (b(e_{13}-ie_{23})+
	  a(e_{31}-ie_{32}))\otimes(c(e_{11}+e_{22})+d(e_{12}-e_{21})+fe_{33})$ \\
\hline $39$ & $S_4\circ Z_2$ & $6$ & \rule[-2ex]{0pt}{5ex}
		$\eta\otimes\overline\eta\otimes\delta$  \\
\hline  $40$ & $S_3\times Z_2$ & $12$ & \rule[-2ex]{0pt}{5ex}
	  $(a\delta+b\varkappa)^{\otimes2}\otimes(c\delta+d\varkappa)$ \\
\hline  $41$ & $S_3\times Z_2$ & $12$ & \rule[-2ex]{0pt}{5ex}
	 $\tau^{\otimes2}\otimes (a\delta+b\varkappa)$ \\
\hline  $42$ &  $A_4\circ Z_3$ & $12$ & \rule[-2ex]{0pt}{5ex}
	 $(ae_{11}+be_{22}+ce_{33})\otimes (ce_{11}+ae_{22}+be_{33})\otimes
	(be_{11}+ce_{22}+ae_{33})$  \\
\hline  $43$ & $S_4\circ_2 S_3$ & $6$ & \rule[-2ex]{0pt}{5ex}
	$(ae_{11}+b(e_{22}+e_{33}))\otimes(ae_{22}+b(e_{11}+e_{33}))\otimes
	(ae_{33}+b(e_{11}+e_{22}))$ \\
\hline  $44$ & $S_4\circ_2 S_3$ & $6$ & \rule[-2ex]{0pt}{5ex}
	$(ae_{23}+be_{32})\otimes(be_{13}+ae_{31})\otimes(ae_{12}+be_{21})$ \\
\hline
\end{longtable}
}
\smallskip
The rest of the article is devoted to the proof of this theorem.
\smallskip

{\bf 5. Semiinvariants for the subgroups of $A$.} Suppose
$w=x\otimes y\otimes z\in M^{\otimes3}$ is a decomposable tensor
whose $G$-orbit is of length $\leq18$. Then its stabilizer $H={\rm
St}_G(w)$ is of order $\geq8$, and so has nontrivial intersection
with~$A$. Therefore, there exists $a\in A$ such that
$axa^{-1}\otimes aya^{-1}\otimes aza^{-1}= x\otimes y\otimes z$,
whence $axa^{-1}$, $aya^{-1}$, and $aza^{-1}$ must be proportional
to $x,y$, and $z$, respectively. So the following task is
reasonable: for each {\em nontrivial} subgroup $K\leq A$ find all
its semiinvariants in $M$, i.e. all $x\in M$ such that $axa^{-1}$
is proportional to $x$, for all $a\in K$. Clearly, then there
exists a homomorphism $\lambda\:\colon\: K\longrightarrow{\mathbb
C}^\ast$ ({\em character}) of $K$ such that $axa^{-1}=\lambda(a)x$
for each $a\in A$. (In the present article the word ``character''
always means one-dimensional (``linear'') character, that is, a
homomorphism to ${\mathbb C}^\ast$, and is used without an
adjective.)

Let $K^\ast={\rm Hom}(K,{\mathbb C}^\ast)$ be the group of
characters of $K$, and for $\lambda\in K^\ast$ let
$$ M_\lambda=\{ x\in M\mid axa^{-1}=\lambda(a)x,\ \forall a\in K\} $$
be the corresponding invariant subspace. Then the sum
$\sum_{\lambda\in K^\ast} M_\lambda$ is always direct. If $K$ is
abelian, then $\bigoplus_{\lambda\in K^\ast}M_\lambda=M$, whereas
for nonabelian $K$ this sum is a proper subspace of~$M$.

Thus, we come to the following question: for each subgroup $1\ne
K\leq A$ find the subspaces $M_\lambda$, $\lambda\in K^\ast$.

It will be convenient to use the same notation for subgroups of
$S_4$ and the corresponding subgroups of $A$. Let
$K_1=Z_2^{(1)},\ldots,K_{10}=S_4$ be the canonical representatives
of the conjugacy classes of the subgroups of~$S_4$. In the
following proposition we list the nontrivial subspaces
$M_\lambda$, for $\lambda\in(K_i)^\ast$. We denote them by
$L_{i,j}$.
\smallskip

{\bf Proposition  5.} {\em 1) The nontrivial subspaces of
semiinvariants for $K_i$ in $M$ are the following:
\begin{itemize}
\item $K_1=Z_2^{(1)}$: \ \ \ $L_{1,1}=\langle e_{11}+e_{22},
e_{12}+e_{21}, e_{13}+e_{23}, e_{31}+e_{32}, e_{33}\rangle$,
$L_{1,2}=\langle e_{11}-e_{22}, e_{12}-e_{21}, e_{13}-e_{23},
e_{31}-e_{32} \rangle$,
\item $K_2=Z_2^{(2)}$: \ \ \ $L_{2,1}=\langle e_{11}, e_{22},
e_{21}, e_{22}, e_{33}\rangle$, $L_{2,2}=\langle e_{13}, e_{23},
e_{31}, e_{32}\rangle$,
\item $K_3=Z_3$: \ \ \ $L_{3,1}=\langle
\delta=e_{11}+e_{22}+e_{33}, e_{12}+e_{23}+e_{31},
e_{13}+e_{21}+e_{32}\rangle$, $L_{3,2}=\langle \eta, e_{12}+\zeta
e_{23}+\overline\zeta e_{31}, e_{13}+\zeta e_{21}+\overline\zeta
e_{32}\rangle$, $L_{3,3}=\langle \overline\eta,
e_{12}+\overline\zeta e_{23}+\zeta e_{31}, e_{13}+\overline\zeta
e_{21}+\zeta e_{32}\rangle$;
\item $K_4=V^{(1)}$: \ \ \ $L_{4,1}=\langle e_{11}+e_{22},
e_{12}+e_{21}, e_{33}\rangle$, $L_{4,2}=\langle e_{11}-e_{22},
e_{12}-e_{21}\rangle$, $L_{4,3}=\langle e_{13}+e_{23},
e_{31}+e_{32}\rangle$, $L_{4,4}=\langle e_{13}-e_{23},
e_{31}-e_{32}\rangle$;
\item $K_5=V^{(2)}$: \ \ \ $L_{5,1}=\langle
e_{11},e_{22},e_{33}\rangle$, $L_{5,2}=\langle
e_{12},e_{21}\rangle$, $L_{5,3}=\langle e_{13},e_{31}\rangle$,
$L_{5,4}=\langle e_{23},e_{32}\rangle$;
\item $K_6=Z_4$: \ \ \ $L_{6,1}=\langle e_{11}+e_{22},
e_{12}-e_{21}, e_{33}\rangle$, $L_{6,2}=\langle e_{12}+e_{21},
e_{11}-e_{22}\rangle$, $L_{6,3}=\langle e_{13}+ie_{23},
e_{31}+ie_{32}\rangle$, $L_{6,4}=\langle e_{13}-ie_{23},
e_{31}-ie_{32}\rangle$, ($i^2=-1$);
\item $K_7=S_3$: \ \ \ $L_{7,1}=\langle \delta, \varkappa\rangle$,
$L_{7,2}=\langle\tau\rangle$,
\item $K_8=D_8$: \ \ \ $L_{8,1}=\langle e_{11}+e_{22},
e_{33}\rangle$, $L_{8,2}=\langle e_{11}-e_{22}\rangle$,
$L_{8,3}=\langle e_{12}+e_{21}\rangle$, $L_{8,4}=\langle
e_{12}-e_{21}\rangle$;
\item $K_9=A_4$: \ \ \ $L_{9,1}=\langle \delta\rangle$,
$L_{9,2}=\langle \eta\rangle$,
$L_{9,3}=\langle\overline\eta\rangle$;
\item $K_{10}=S_4$: \ \ \ $L_{10,1}=\langle\delta\rangle$.
\end{itemize}

2) Denote by $\chi_{ij}$ the character of $K_i$ corresponding
to~$L_{ij}$. Then always $\chi_{i,1}=1$. The following relations
hold:
$$\chi_{s,2}^2=1,\ s=1,2,7$$
(in these cases the group of characters $K_s^\ast\cong Z_2$);
$$\chi_{s,2}^2=\chi_{s,3}\,,\qquad \chi_{s,2}^3=1\,,\qquad
  K_s^\ast\cong Z_3\,,\qquad    s=3,9;$$
$$\chi_{s,2}^2=\chi_{s,3}^2=\chi_{s,4}^2=\chi_{s,2}\chi_{s,3}\chi_{s,4}=1\,,
   \qquad    K_s^\ast\cong Z_2\times Z_2\,,\qquad s=4,5,8\,,$$
$$\chi_{6,3}^2=\chi_{6,2}\,,\qquad \chi_{6,3}^3=\chi_{6,4}\,,\qquad
   \chi_{6,3}^4=1\,,    \qquad K_6^\ast\cong Z_4.$$
}
\smallskip

{\bf Sketch of a proof.} 1) This is a rather direct computation.
For example consider  $K_6=Z_4$ (the hardest case). It is easy to
see that the subgroup of $A$, corresponding to standard $Z_4\leq
S_4$, is generated by the matrix
$$ g=\begin{pmatrix} 0 & 1 & 0 \\ -1 & 0 & 0 \\ 0 & 0 & 1 \end{pmatrix}
= e_{12}-e_{21}+e_{33}\,.$$

Obviously, $g^{-1}=-e_{12}+e_{21}+e_{33}$. Hence it is easy to
calculate that the conjugation by $g$ acts on matrix units by
$$ e_{11}\leftrightarrow e_{22}\,,\quad e_{12}\mapsto -e_{21}\,,\quad
   e_{21}\mapsto -e_{12}\,,\quad   e_{33}\mapsto e_{33}\,,$$
$$ e_{13}\mapsto -e_{23}\,,\quad e_{23}\mapsto e_{13}\,,\quad e_{31}\mapsto
   -e_{32}\,, \quad e_{32}\mapsto e_{31}\,. $$
Now it is easy to see that the subspaces of matrices such that the
conjugation by $g$ multiplies the matrix by $1$, $-1$, $i$, or
$-i$, are $L_{6,1}$, $L_{6,2}$, $L_{6,3}$, and $L_{6,4}$,
respectively.

To treat the other cases and to prove statement 2) is left to the
reader. (For nonabelian groups it is useful to note that the
commutator subgroup must act on any semiinvariant identically.)
\hfill $\square$
\smallskip

{\bf 6. Some invariance conditions.} In this section we collected
several {\em general} statements concerning invariance properties
of decomposable tensors under various subgroups of $S_4\times
S_3$. (Particular subgroups will be considered later).

We identify $G$ with $S_4\times S_3$ and write an element of $G$
as a pair of permutations.

Below $w$ usually denotes a decomposable tensor $w=x\otimes
y\otimes z\in M^{\otimes3}$. The next statement is obvious.
\smallskip

{\bf Proposition 6.} {\em Let $K\leq S_4$. Then $w$ is invariant
under $K\times1\leq G$ if and only if $x$, $y$, and $z$ are
semiinvariants for $K$ belonging to characters $\mu_{1,2,3}\in
K^\ast$ such that $\mu_1\mu_2\mu_3=1$. \hfill $\square$ }
\smallskip

In the rest of the article we often use the following simple
statement.
\smallskip

{\bf Lemma 7.} {\em Let $a\in A$ (or, more generally, $a$ is any
orthogonal $3\times3$ matrix). Then the transformations $x\mapsto
axa^{-1}$ and $x\mapsto x^t$ of $M$ commute, so that
$(axa^{-1})^t=ax^ta^{-1}$. }
\smallskip

{\bf Proof.} We have $(axa^{-1})^t=(a^{-1})^tx^ta^t=ax^ta^{-1}$ as
$a^t=a^{-1}$, since $a$ is orthogonal. \hfill $\square$
\smallskip

{\bf Corollary 8.} {\em  If $K\leq S_4$, and if $x\in M$ is a
semiinvariant for $K$ belonging to a character $\mu\in K^\ast$,
then $x^t$ is also a semiinvariant belonging to the same
character. Thus, all the subspaces $L_{i,j}$ of Proposition 5 are
invariant under taking transpose (which, anyway, is visible
immediate from the explicit form of the spaces $L_{i,j}$). }
\smallskip

For an element $h\in S_4$ the statement of the lemma may be
written as $(hx)^t=hx^t$, if we mean by $hx$ the image of $x$
under $h$, i.e. $hx=\widehat hx\widehat h^{-1}$, where $\widehat
h=\gamma(h)$.
\smallskip

{\bf Proposition 9.} {\em 1) A decomposable tensor $w$ is
invariant under $1\times Z_3\leq G$ if and only if
$w=x^{\otimes3}$, for some $x\in M$.

2) $w$ is invariant under $1\times Z_2$ $\Longleftrightarrow$
$w=x\otimes x^t\otimes y$, and $y=y^t$.

3) $w$ is invariant under $1\times S_3$ $\Longleftrightarrow$
$w=x^{\otimes3}$, $x^t=x$.

4) $w$ is invariant under $K\times Z_3$ $\Longleftrightarrow$
$w=x^{\otimes3}$, where $x$ is a semiinvariant for $K$ belonging
to a character $\mu\in K^\ast$ such that $\mu^3=1$.

5) $w$ is invariant under $K\times Z_2$ $\Longleftrightarrow$
$w=x\otimes x^t \otimes y$, $y=y^t$, and $x\in M_\lambda$, $y\in
M_\mu$, $\lambda^2\mu=1$.

6) $w$ is invariant under $K\times S_3$ $\Longleftrightarrow$
$w=x^{\otimes3}$, $x=x^t$, $x\in M_\mu$, $\mu^3=1$. }
\smallskip

{\bf Proof.} 1) We have $1\times Z_3=\langle\sigma\rangle$, where
the element $\sigma=(1,(123))\in G$ acts by $\sigma\colon x\otimes
y\otimes z\mapsto z\otimes x\otimes y$. Obviously, any tensor of
the form $x\otimes x\otimes x$ is $\sigma$-invariant. Conversely,
suppose that $\sigma w=w$. Then $z=\lambda_1x$, $x=\lambda_2y$,
and $y=\lambda_3z$ for some $\lambda_{1,2,3}\in{\mathbb C}$, and
$\lambda_1\lambda_2\lambda_3=1$. Hence $y$ and $z$ are
proportional to $x$ with nonzero coefficients, and therefore
$w=mx^{\otimes3}$, $m\ne0$, whence $w=(x')^{\otimes3}$,
$x'=m^{1/3}x$.

2) We have similarly $1\times Z_2=\langle\rho\rangle$, where
$\rho\colon x \otimes y\otimes z\mapsto y^t\otimes x^t\otimes
z^t$. So a tensor of the form $x\otimes x^t\otimes z$, where
$z=z^t$ is symmetric, is invariant under $\rho$. Conversely, the
equality $\rho w=w$ implies $y=\lambda x^t$, so $w=x\otimes
x^t\otimes z'$ for appropriate $z'$. Hence $\rho w=(x^t)^t \otimes
x^t\otimes (z')^t$, and equality $\rho w=w$ implies $(z')^t=z'$.
It remains to denote $z'$ by $y$.

3) Obviously, $\rho(x^{\otimes3})=(x^t)^{\otimes3}$ for any $x\in
M$. So the tensor $x^{\otimes3}$ with symmetric $x=x^t$ is
invariant under $1\times S_3=B=\langle\sigma,\rho\rangle$.
Conversely, suppose $w$ is invariant under~$B$. The invariance
under $\sigma$ implies $w=x^{\otimes3}$. 
Now the invariance under
$\rho$ gives $(x^t)^{\otimes3}=x^{\otimes3}$, whence $x^t=\lambda
x$, $\lambda^3=1$. 
But taking transpose can not multiply a matrix
by a nontrivial cubic root of~$1$.

The statements 4), 5), and 6) easily follow from Proposition 6 and
statements 1), 2), and 3), respectively. \hfill $\square$
\smallskip

In order to consider tensors invariant under groups of the form
$K\circ Z_2$ or $K\circ S_3$ with common quotient group $\cong
Z_2$, we need to know a condition of invariance under elements of
the form $(h,(12))\in S_4\times S_3$, where $h\in S_4$ is of order
$2$ or $4$.
\smallskip

{\bf Proposition 10.} {\em 1) Let $h\in S_4$ be an element of
order $2$, and $g=(h,(12))\in G$. Then $w=x\otimes y\otimes z$ is
invariant under $g$ if and only if $w$ is of the form $x\otimes
Rx\otimes z$, where $R$ is the transformation of $M$ defined by
$Ry=(hy)^t$, and $z$ satisfies $Rz=z$.

2) Suppose $h$ is of order $1$, $2$, or $4$. Then $x^{\otimes3}$
is invariant under $g$ $\Longleftrightarrow$ $Rx=x$. }
\smallskip

{\bf Proof.}  1)  For an arbitrary decomposable tensor we have
\begin{eqnarray*} g(x\otimes y\otimes z) &=& (1,(12))((h,1)
(x\otimes y\otimes z))=(1,(12))(hx\otimes hy\otimes hz) \\
&=& (hy)^t\otimes(hx)^t\otimes(hz)^t=Ry\otimes Rx\otimes Rz.
\end{eqnarray*}
So, if $w$ is $g$-invariant, then $y$ is proportional to $Rx$,
that is, $w$ is of the form $x\otimes Rx\otimes z'$.

Note that $R$ is an involutive transformation, since the
transformations $v\mapsto v^t$ and $v\mapsto hv$ commute and both
are of order~$2$. Therefore, $g(x\otimes Rx\otimes z')=R^2x\otimes
Rx\otimes Rz'=x\otimes Rx \otimes Rz'$. So $gw=w$ implies
$Rz'=z'$. (And vice versa, if $Rz=z$, then $x\otimes Rx\otimes z$
is clearly $g$-invariant.)

2) Obviously, $g(x^{\otimes3})=(Rx)^{\otimes3}$. So the
$g$-invariance of $x^{\otimes3}$ implies $Rx=\mu x$, where
$\mu^3=1$. Since $h$ is of order $1$, $2$, or $4$, the
transformation $R$ is of order $1$, $2$, or $4$ also, and so can
not multiply $x$ by a nontrivial cubic root of~$1$. \hfill
$\square$
\smallskip

Let $X$ and $Y$ be the subgroups consisting of all elements of the
form $(\pi,\pi)$, where $\pi$ runs over $Z_3$ for $X$ and over
$S_3$ for $Y$. That is $X=\langle g_5\rangle_3$ and $Y=\langle
g_2,g_5\rangle$ in the notation of Proposition~3. Let us find the
general form of the decomposable tensors invariant under $X$ or
$Y$.

It is clear that $(\pi,\pi)(x\otimes y\otimes z)$ is obtained from
$x\otimes y\otimes z$ first by the componentwise action of $\pi$,
and then by the permutation of the factors according to $\pi$. And
in addition, if $\pi$ is odd, then we must transpose each factor.
In particular,
$$ ((12),(12))(x\otimes y\otimes z)=((12)y^t, (12)x^t,(12)z^t),$$
\vspace{-1.0ex}
$$ ((123),(123))(x\otimes y\otimes z)=((123)z, (123)x,(123)y). $$

{\bf Proposition 11.} {\em The decomposable tensors invariant
under $X=\langle g_5\rangle_3$ are precisely all the tensors of
the form $x\otimes(123)x\otimes(132)x$, $x\in M$, or,
equivalently, all the tensors of the form
$$(123)x\otimes(132)x\otimes x.\eqno (1)$$
The tensors invariant under $Y$ are the tensors of the form (1)
with $x$ satisfying condition $(12)x^t=x$. }
\smallskip

{\bf Proof.} The condition that $x\otimes y\otimes z$ is invariant
under $((123),(123))$ is clearly equivalent to relations
$(123)z=\lambda_1x$, $(123)x=\lambda_2y$, $(123)y=\lambda_3z$,
with $\lambda_1\lambda_2\lambda_3=1$. Therefore, $y$ is
proportional to $(123)x$ and $z$ to $(123)^{-1}x=(132)x$. Whence
$x\otimes y\otimes z$ is proportional to
$x\otimes(123)x\otimes(132)x$.

Note that any tensor proportional to a tensor of the latter form
is of this form too (it is sufficient to divide $x$ by the cubic
root of the proportion coefficient). Conversely, it is clear that
any tensor of the form $x\otimes (123)x\otimes(132)x$ (or,
equivalently, of the form $(123)x\otimes(132)x \otimes x$) is
invariant under $((123),(123))$.

The tensors invariant under $Y$ also, clearly, are of the
form~(1). Moreover, it follows from Proposition 10, 1), that they
satisfy condition $(12)x^t=x$.

It remains to show the converse, that is the tensors of the form
(1) with additional restriction $(12)x^t=x$ are invariant under
$Y$. To do this it suffices to show that they are invariant under
$((12),(12))$, and for this, in turn, to check that $x_1=(123)x$
and $x_2=(132)x$ differ by an action of $R\colon v\mapsto(12)v^t$.
Indeed,
$$ Rx_1=(12)((123)x)^t=(12)(123)x^t=(12)(123)((12)x)=((12)(123)(12))x=
  (132)x=x_2$$
(we use here the relation $x^t=(12)x$). As $R$ is involutive, we
have also $Rx_2=x_1$. \hfill $\square$ \smallskip

{\bf 7. Beginning of the proof of Theorem 4.} Now we start to
prove Theorem~4. In this section we discuss the easier part of the
theorem (``the orbit of the tensor $w_i(a,b,\ldots)$ has length
$l_i$, for almost all $(a,b,\ldots)$'', and the more hard (but, in
fact, easy enough also!) part (``any orbit is generated by a
tensor $w_i(a,b,\ldots)$'') consider in the next sections.
\smallskip

First of all, for each $i=1,\ldots,44$ we need to check that
$w_i(a,b,\ldots)$ is invariant under~$H_i$. By Proposition 3, any
subgroup $H=H_i$ can be represented as $(C\times D)R$, where $R$
is one of the groups $1$, $\langle g_i\rangle$, $i=1,\ldots,5$, or
$\langle g_2,g_5\rangle$. The invariance of $w=x\otimes y\otimes
z$ under $C$ can be easily checked by means of Proposition~6,
under $D$ with Proposition~9, and under $R$ with Propositions 10
and~11.

For example consider the case $i=38$, where the tensor
$w_i(a,b,\ldots)$ looks probably most complex. We have
$w_{38}(a,b,c,d,f)=x\otimes y\otimes z$, where
$$ x=a(e_{13}+ie_{23})+b(e_{31}+ie_{32})\,, \qquad y=b(e_{13}-ie_{23})+
   a(e_{31}-ie_{32})\,,$$
$$ z=c(e_{11}+e_{22})+d(e_{12}-e_{21})+fe_{33}; $$
$$H_{38}=D_8\circ_3 Z_2=(Z_4\times1)\leftthreetimes\langle g_2\rangle_2\,. $$
By Proposition 5, $x\in L_{6,3}$, $y\in L_{6,4}$, $z\in L_{6,1}$,
and $\chi_{6,3}\chi_{6,4}\chi_{6,1}=1$, so $w$ is invariant under
$Z_4\times1$. Further, the condition for invariance under
$g_2=((12),(12))$ is that $y$ is proportional to $Rx$, and $z=Rz$,
where $Rv=((12)v)^t$.

Taking into account the remark in the end of Section 3 (that $\pi
e_{ij}= \widehat\pi e_{ij}\widehat\pi^{-1}=e_{\pi i,\pi j}$), we
have
\begin{eqnarray*}
Rx &=& ((12)x)^t=((12)[a(e_{13}+ie_{23})+b(e_{31}+ie_{32})])^t \\
&=& (a(e_{23}+ie_{13})+b(e_{32}+ie_{31}))^t=a(e_{32}+ie_{31})+
b(e_{23}+ie_{13})=iy,
\end{eqnarray*}
and also
\begin{eqnarray*}
Rz &=& ((12)[c(e_{11}+e_{22})+d(e_{12}-e_{21})+fe_{33}])^t=
(c(e_{22}+e_{11})+d(e_{21}-e_{12})+fe_{33})^t \\
 &=& c(e_{11}+e_{22})+d(e_{12}-e_{21})+fe_{33}=z\,,
\end{eqnarray*}
which proves that $w$ is invariant.   The detailed calculations
for other cases are left to the reader.
\smallskip

Next, we should show that for almost all sets of parameters
$x=(a,b,\ldots)$ the orbit of tensor $w_i(x)$ is of length $l_i$,
or equivalently, the stabilizer coincides with $H_i$ (the reader
can easily see that $l_i=|G:H_i|$ in all the cases listed in the
table), an that the exceptional $x$ form a proper subvariety in
${\mathbb C}^{s_i}$ (where $s_i$ is the number of parameters
$a,b,\ldots$, for given~$i$).

It is clear that $\varphi_i\colon x\mapsto w_i(x)$ is a polynomial
map from ${\mathbb C}^{s_i}$ to $M^{\otimes3}$. The set of all
(not necessary decomposable) tensors invariant under given $g\in
G$ is a subspace of $M^{\otimes3}$. So its $\varphi_i$-preimage is
a Zariski closed subset of ${\mathbb C}^{s_i}$. That is, the set
of all $x$ such that $w_i(x)$ is $g$-invariant, is closed. The set
of all ``exceptional'' $x$ is the union of the latter sets over
all $g\in G\setminus H_i$ and so is closed also. To show that this
is a proper subvariety it suffices to produce, for each $1\leq
i\leq44$, a set of parameters for which the length of the orbit is
$\geq l_i$. This usually can be done orally, without many
computations. As an example consider again $i=38$.

Take $w$ general enough, an at the same time simple enough. Say,
in the case under consideration we can take
$$ w=(e_{13}+ie_{23})\otimes(e_{31}-ie_{32})\otimes e_{33}\,.$$

Let ${\mathcal O}$ be its orbit. It is clear that for each tensor
$x\otimes y\otimes z\in{\mathcal O}$ one of the vectors $x$, $y$,
$z$ is of the form $\pm e_{ik}\pm ie_{jk}$ (do not confuse here
subscript $i$ with the imaginary unit $i=\sqrt{-1}$!), the other
of the form $\pm e_{ki}\pm ie_{kj}$, and the remaining third of
the form $e_{kk}$. This vector $e_{kk}$ may be in the place of
$x$, $y$, or $z$, and this gives us a partition of ${\mathcal O}$
into three disjoint subsets ${\mathcal O}={\mathcal O}_1
\sqcup{\mathcal O}_2\sqcup{\mathcal O}_3$, which are clearly of
the same cardinality (it is sufficient to consider the action of
$1\times Z_3\leq G$). We can assume that ${\mathcal O}_1$ is the
subset corresponding to $z=e_{kk}$.

Next, ${\mathcal O}_1$ splits into three subsets ${\mathcal O}_1=
{\mathcal O}_1'\sqcup{\mathcal O}_1''\sqcup{\mathcal O}_1'''$,
corresponding to the cases $k=1,2,3$ (consider the action of
$Z_3\times1$). Finally, ${\mathcal O}_1'''$ contains at least two
points $w$ and $w'\ne w$,
$$ w'=(e_{13}-ie_{23})\otimes(e_{31}+ie_{32})\otimes e_{33}$$
($w'=(h,1)w$, where $h=(13)(24)$, which corresponds to $\widehat
h={\rm diag} (-1,1,-1)\in A$). Thus, we see that $|{\mathcal
O}|=3|{\mathcal O}_1|= 9|{\mathcal O}_1'''|\geq18$, as required.
\medskip

{\bf 8. $G$-orbits. The case of $1\times Z_3\leq H$.} Now we begin
to prove the more difficult part of Theorem~4. Observe first of
all that for a given $G$-orbit ${\mathcal O}$ of length $\leq18$
on decomposable tensors the stabilizers of points of ${\mathcal
O}$ form a conjugacy class of subgroups of $G$, so there exists a
point $w\in{\mathcal O}$ whose stabilizer is a standard
representative of conjugacy class of subgroups, that is one of the
subgroups listed in Proposition~2. We call this subgroup, denote
it $H$, {\em the stabilizer of the orbit}. So, we should find all
$H$-invariant decomposable tensors, for each of these standard
subgroups $H$.

However, note that the set of all $G$-orbits whose stabilizer is
$H$ {\em is not}, in general, in a bijection with the set of all
$H$-invariant decomposable tensors, for two reasons. First, it may
be that for a given $H$-invariant decomposable tensor $w$ its
$G$-stabilizer ${\rm St}_G(w)$ strictly contains $H$. Then the
length of its orbit is actually $<|G:H|$. In particular, it may
happen that the orbits with a given stabilizer $H$ do not exist at
all. Second, it may be that the normalizer $N_G(H)$ strictly
contains $H$: $N_G(H)>H$. Then the orbit $Gw$ contains {\em
several} tensors, whose stabilizers are equal to $H$ (namely, all
the tensors of the form $gw$, with $g\in N_G(H)/H$).

Taking these remarks into account, the following terminology will
be used. We say that Theorem 4 is {\em true for a standard
subgroup $H$}, if any orbit ${\mathcal O}$, whose stabilizer is
$H$, is generated by a tensor of the form $w=w_i(a,b,\ldots)$,
where $1\leq i\leq44$ is one of the indices such that $H_i=H$. In
particular, the theorem will be true for $H$ if there exist no
orbits with stabilizer $H$ and in the table of Theorem 4 there are
no rows with $H_i=H$ at all.

Obviously, to prove Theorem 4 it is sufficient to prove it for all
standard subgroups $H$. It is clear that if each decomposable
tensor whose stabilizer is $H$, is of the form $w_i(a,b,\ldots)$
with $H_i=H$, then the theorem is true for $H_i$; but the converse
is not true, in general.

Note that the majority of the subgroups listed in Proposition 2
contains $1\times Z_3$ as a subgroup. In this section we prove the
theorem for all such subgroups.

Consider $H=Z_2^{(1)}\times S_3$. By Proposition 9.6), any
$H$-invariant tensor is of the form $x^{\otimes3}$, where $x$ is a
semiinvariant belonging to a character $\mu$ of $Z_2^{(1)}$ such
that $\mu^3=1$. Moreover, $x$ must be symmetric. But there are no
such $\mu$, except for $\mu=1$. So $x$ is an invariant for
$Z_2^{(1)}$, $x\in L_{1,1}=\langle e_{11}+e_{22}, e_{12}+e_{21},
e_{13}+e_{23}, e_{31}+e_{32}, e_{33}\rangle$. In particular, $x$
is necessary symmetric. Therefore any $H$-invariant tensor is of
the form $w_1(a,b,c,d,f)$, and the theorem is proved for this~$H$.

The same argument applies for $H=Z_2^{(2)}\times S_3$,
$V^{(1)}\times S_3$, $V^{(2)}\times S_3$, $S_3\times S_3$,
$D_8\times S_3$, $S_4\times S_3$. On the other hand, there exists
no orbits whose stabilizer is one of the groups $K\times Z_3$,
$K=V^{(1)}, V^{(2)}$, $S_3$, $D_8$, $S_4$. For a $K\times
Z_3$-invariant tensor must be of the form $x^{\otimes3}$, where
$x$ is an invariant for~$K$. But in all these cases all the
$K$-invariants in $M$ are symmetric, so in fact $x^{\otimes3}$ is
invariant under $K\times S_3$.
\medskip

Next consider three groups $Z_3\times Z_3$, $Z_3\times S_3$, and
$S_3\circ S_3$ (note that the latter two contain $Z_3\times Z_3$).
A tensor invariant under $Z_3\times Z_3$ is $x^{\otimes3}$, where
$x\in L_{3,1}$, $L_{3,2}$, or $L_{3,3}$. By Proposition~3,
$S_3\circ S_3=(Z_3\times Z_3)\leftthreetimes \langle
g_2\rangle_2$, $g_2=((12),(12))$. We have $g_2(x\otimes y\otimes
z)=Ry\otimes Rx\otimes Rz$, where $Rv=((12)v)^t$. It is easy to
see that for $v\in L_{3,1}$ we have $Rv=v$. Moreover, $(12)\in
S_4$ (keep in mind that here ``$(12)\in S_4$'' actually means the
element of $A$, or $A\times1$, corresponding to $(12)\in S_4)$,
and therefore $R$, interchanges $L_{3,2}$ with $L_{3,3}$. Hence
the tensor $x^{\otimes3}$ is $g_2$-invariant when $x\in L_{3,1}$
and is not $g_2$-invariant when $x\in L_{3,2}, L_{3,3}$. So, the
$S_3\circ S_3$-invariant tensors are precisely the tensors of the
form $w_{11}(a,b,c)$, and the theorem is true for $S_3\circ S_3$.
And since $((12),1)$ interchanges the tensors $x^{\otimes3}$,
where $x\in L_{3,2}$, with such tensors with $x\in L_{3,3}$, then
any orbit whose stabilizer is $Z_3\times Z_3$ has a representative
$x^{\otimes3}$ with $x\in L_{3,2}$, i.e., of the form
$w_8(a,b,c)$, which proves the theorem for $Z_3\times Z_3$.

Finally assume that $x^{\otimes3}$ is invariant under $Z_3\times
S_3$. Then $x\in L_{3,1}$, $L_{3,2}$, or $L_{3,3}$, and $x$ is
symmetric. If $x\in L_{3,1}$, then $x$ is invariant under $(12)\in
S_4$ and so $x^{\otimes3}$ is invariant under $S_3\times S_3$, a
contradiction. And since $(12)\in S_4$ normalizes $Z_3\times S_3$
and interchanges $L_{3,2}$ with $L_{3,3}$, we see that any orbit
whose stabilizer is $Z_3\times S_3$ contains a tensor of the form
$x^{\otimes3}$, where $x\in L_{3,2}$ and is symmetric, that is a
tensor of the form $w_{10}(a,b)$, which proves the theorem for
$H=Z_3\times S_3$.
\medskip

Next consider $A_4\times Z_3$, $A_4\times S_3$, and $S_4\circ_1
S_3$. The tensors, invariant under $A_4\times Z_3$, are
$x^{\otimes3}$, where $x\in L_{9,1}$, $L_{9,2}$, or $L_{9,3}$.
That is they are multiples of $\delta^{\otimes3}$,
$\eta^{\otimes3}$, or $\overline\eta^{\otimes3}$, respectively.
The tensor $a\delta^{\otimes3}$ is invariant under $S_4\times
S_3$, that is it is a one-point orbit. The pair
$\{a\eta^{\otimes3}, a\overline\eta^{\otimes3}\}$ is an orbit with
stabilizer $A_4\times S_3$. So an orbit whose stabilizer is
$H=A_4\times S_3$ contains $a\eta^{\otimes3}=w_6(a)$, and there
exist no orbits with stabilizer $H=A_4\times Z_3$ or $S_4\circ_1
S_3$.

Next consider group $Z_4\times Z_3$ and two groups containing it,
namely $Z_4\times S_3$ and $D_8\circ_3 S_3$. If $x^{\otimes3}$ is
invariant under $Z_4\times Z_3$, then $x$ is invariant under
$Z_4$, whence $x\in L_{6,1}= \langle e_{11}+e_{22}, e_{12}-e_{21},
e_{33}\rangle$. But any $x\in L_{6,1}$ is invariant under $R\colon
x\mapsto((12)x)^t$. So $x^{\otimes3}$ is invariant under
$g_2=((12),(12))$, and therefore under $(Z_4\times Z_3)
\leftthreetimes\langle g_2\rangle_2=D_8\circ_3 S_3$. Thus, in the
case $H=D_8\circ_3 S_3$ we have $w=w_{12}(a,b,c)$, and the cases
$H=Z_4\times Z_3$ or $Z_4\times S_3$ are impossible (in the last
case ${\rm St}_G(w)$ would contain $\langle Z_4\times S_3,\
g_2\rangle=D_8\times S_3$).

The cases $D_8\circ_1S_3$ and $D_8\circ_2S_3$ are impossible.
Indeed, this subgroups contain $V^{(1)}\times Z_3$ and
$V^{(2)}\times Z_3$, respectively. But, as we have seen earlier,
any $x^{\otimes3}$ that is invariant under these subgroups is
invariant also under $1\times S_3$, so ${\rm St}_G(w)$ is strictly
larger then $D_8\circ_iS_3$, $i=1,2$.
\medskip

Next we consider, in a uniform way, three groups
$Q_1=V^{(1)}\circ_1S_3$, $Q_2=V^{(1)}\circ_2S_3$, and
$Q_3=V^{(2)}\circ S_3$. These groups can be represented as
$Q_i=(P_i\times Z_3)\leftthreetimes\langle (h_i,(12))\rangle_2$,
where $P_{1,2,3}=Z_2^{(1)}$, $Z_2^{(2)}$, $Z_2^{(2)}$,
respectively, and $h_{1,2,3}=(12)(34)$, $(12)$, $(13)(24)$. So the
set of $Q_i$-invariant decomposable tensors coincides with the set
of tensors $x^{\otimes3}$, where $x$ is invariant under both $P_i$
and $R_i$, where $R_i\colon x\mapsto (h_ix)^t$ (see Proposition
10.2)).

The spaces of $P_i$-invariants in $M$ are $N_1=L_{1,1}$ and
$N_2=N_3=L_{2,1}$. The spaces of $R_i$-invariants are
$$ N'_1=\langle e_{11},e_{22}, e_{12}+e_{21},e_{33}, e_{13}-e_{31},
   e_{23}-e_{32}\rangle\,, $$
$$N'_2=\langle e_{11}+e_{22}, e_{12},e_{21},e_{33}, e_{13}+e_{32},
	e_{23}+e_{31}\rangle\,,$$
$$ N'_3=\langle e_{11},e_{22}, e_{33}, e_{12}-e_{21}, e_{13}+e_{31},
   e_{23}-e_{32}\rangle\,. $$
The intersections $U_i=N_i\cap N'_i$ are
$$ U_1=\langle e_{11}+e_{22}, e_{12}+e_{21},e_{33}, e_{13}-e_{31}+e_{23}-
   e_{32}\rangle\,, $$
$$ U_2=\langle e_{11}+e_{22}, e_{12},e_{21},e_{33}\rangle\,,\qquad
   U_3=\langle e_{11},e_{22}, e_{33}, e_{12}-e_{21}\rangle\,. $$
The tensors $x^{\otimes3}$ with $x\in U_{1,2,3}$ are precisely the
tensors of the form $w_i(a,b,\ldots)$ with $i=13,14,15$, which
proves the theorem for the three considered groups.

It remains to show that the case $H=Z_4\circ S_3$ is impossible.
We have $H=(Z_2^{(2)}\times Z_3)\langle((1324),(12))\rangle_4$.
The tensors invariant under $Z_2^{(2)}\times Z_3$ are
$x^{\otimes3}$, $x\in N=\langle
e_{11},e_{22},e_{33},e_{12},e_{21}\rangle$. Find the invariants in
$N$ of the transformation $R\colon x\mapsto((1324)x)^t$. This $R$
can be written as $R(x)=(\widehat hx\widehat h^{-1})^t$, where
$\widehat h=e_{12}-e_{21}+e_{33}$ is the element of $A$
corresponding to $(1324)\in S_4$. Trivially, $\widehat
h^{-1}=-e_{12}+e_{21}+e_{33}$. It is easy to calculate that $R$
transforms basis elements of $N$ as $e_{11} \leftrightarrow
e_{22}$, $e_{33}\mapsto e_{33}$, $e_{12}\mapsto -e_{12}$,
$e_{21}\mapsto -e_{21}$. So the $R$-invariant elements of $N$ are
in $\langle e_{11}+e_{22}, e_{33}\rangle$. But for these $x$ the
tensor $x^{\otimes3}$ is invariant under $D_8\times S_3$, a
contradiction.

Thus, the theorem is true in all the cases where $1\times Z_3\leq
H$.
\medskip

{\bf 9. The orbits whose stabilizer is a 2-group.} In this section
we list the $G$-orbits (of length $\leq18$) on decomposable
tensors, whose stabilizer is a $2$-group. Obviously, in such a
case the length of the orbit is $9$ or $18$, and the order of the
stabilizer is either $16$ or $8$, respectively. The group listed
in Proposition 2 that are 2-groups are $V^{(1)}\times Z_2$,
$V^{(2)}\times Z_2$, $Z_4\times Z_2$, $D_8\times1$,  $D_8\times
Z_2$, and the three groups $D_8\circ_i Z_2$, $i=1,2,3$.

First consider $H=D_8\times1$. Suppose $w=x\otimes y\otimes z$ is
invariant under $H$. Then $x\in L_{8,i}$, $y\in L_{8,j}$, and
$z\in L_{8,k}$. We shall say in such a case that $w$ is {\em of
type $(i,j,k)$}. The invariance condition (Proposition 6) implies
$\chi_{8,i}\chi_{8,j}\chi_{8,k}=1$. Taking into account that the
group of characters $D_8^\ast\cong Z_2^2$, we see that $\{i,j,k\}$
is one of $\{1,1,1\}$, $\{2,2,1\}$, $\{3,3,1\}$, $\{4,4,1\}$, or
$\{2,3,4\}$ as a multiset.

The subgroup $1\times S_3$ normalizes $D_8\times 1$ and so acts on
the set of  $D_8\times1$-invariant tensors. It is easy to see that
if $w=x\otimes y\otimes z$ is of type $(i,j,k)$ and
$w'=(1\times\pi)w=x'\otimes y'\otimes z'$, where $\pi\in S_3$,
then $w'$ is of type $(i',j',k')$, where $(i',j',k')$ is obtained
from $(i,j,k)$ by permutation~$\pi$. So an orbit whose stabilizer
is $D_8\times1$ contains an element of one of types $(1,1,1)$,
$(2,2,1)$, $(3,3,1)$, $(4,4,1)$, or $(2,3,4)$.

The elements of type $(1,1,1)$ are precisely the tensors of the
form $x\otimes y\otimes z$ with $x,y,z\in L_{8,1}=\langle
e_{11}+e_{22},e_{33} \rangle$. The elements of types $(2,2,1)$,
$(3,3,1)$, and $(4,4,1)$ are the tensors of the form
$(e_{11}-e_{22})^{\otimes2}\otimes z$,
$(e_{12}+e_{21})^{\otimes2}\otimes z$, and
$(e_{12}-e_{21})^{\otimes2} \otimes z$ (since the subspaces
$L_{8,i}$ with $i=2,3,4$ are of dimension one). Since $z^t=z$,
Proposition 9.2 implies that these elements are invariant under
$(1,(12))$, so $w$ is invariant under $D_8\times Z_2$, a
contradiction.

Finally, the elements of type $(2,3,4)$ are proportional to
$(e_{11}-e_{22})\otimes (e_{12}+e_{21})\otimes(e_{12}-e_{21})$.

Thus, either $w=x\otimes y\otimes z$ where $x,y,z\in L_{8,1}$ or
$w=a(e_{11}-e_{22})\otimes (e_{12}+e_{21})\otimes(e_{12}-e_{21})$,
$a\in{\mathbb C}^\ast$.

An element of the former of these two forms is
$w_{16}(a,\ldots,g)$, and of the latter $w_{17}(a)$. This proves
the theorem for $D_8\times1$.

Next consider $H=P\times Z_2$, where $P=V^{(1)},V^{(2)},Z_4$, or
$D_8$. Similarly to the case $D_8\times1$, if $w=x\otimes y\otimes
z$ is a $P\times1$-invariant decomposable tensor, then define its
{\em type} as $(l,m,n)$, where $l,m,n$ are such that $x\in
L_{j,l}$, $y\in L_{j,m}$, $z\in L_{j,n}$, and $j$ is the number
such that $K_j=P$ (in the notation of Proposition~5). The
invariance condition gives $\chi_{j,l}\chi_{j,m} \chi_{j,n}=1$.

If $w$ is $P\times Z_2$-invariant, then it is of the form
$x\otimes x^t \otimes z$. As the spaces $L_{p,q}$ are all
invariant under transpose, the type of the latter tensor is
$(l,l,n)$. Hence $\chi_{j,l}^2\chi_{j,n}=1$. Note that in the
cases $P=V^{(1)},V^{(2)},D_8$ we have $P^\ast\cong Z_2^2$, so the
relation $\chi_{j,l}^2\chi_{j,n}=1$ implies $\chi_{j,n}=1$, that
is $n=1$, and the type of $w$ is $(l,l,1)$, $l=1,\ldots,4$. In the
case $P=Z_4$ the type is one of $(1,1,1)$, $(2,2,1)$, $(3,3,2)$,
or $(4,4,2)$.

In fact, these types are sometimes equivalent, in the following
sense. We say that two types $(l,m,n)$ and $(l',m',n')$ are {\em
equivalent}, if every $G$-orbit containing an element of one of
these types necessary contains an element of the other type also.

Let $N=N_{S_4}(P)$ be the normalizer of $P$ in $S_4$. Then $N$
permutes the subspaces $L_{j,i}$, and so acts on $\{1,2,3,4\}$.
The orbits of this action are $\{1\},\{2\},\{3\},\{4\}$ when
$P=D_8$, $\{1\},\{2\},\{3,4\}$ when $P=V^{(1)}$ or $P=Z_4$, and
$\{1\},\{2,3,4\}$ when $P=V^{(2)}$.

Obviously, $(h,1)$ normalizes $P\times Z_2$ if $h\in N$ and so
acts on the set of $P\times Z_2$-invariant decomposable tensors.
Also, if $w$ is of type $(l,m,n)$, then $(h,1)w$ is of type
$(h(l),h(m),h(n))$. It follows that if two types are in the same
orbit with respect to the componentwise action of $N$, then they
are equivalent. Therefore, for $P=V^{(1)}$ any orbit, containing a
$P\times Z_2$-invariant decomposable tensor, contains a tensor of
one of types $(1,1,1)$, $(2,2,1)$, or $(3,3,1)$. When $P=V^{(2)}$,
such an orbit contains a tensor of one of types $(1,1,1)$ or
$(2,2,1)$, and when $P=Z_4$ of types $(1,1,1)$, $(2,2,1)$, or
$(3,3,2)$. It remains to write explicitly for each type $(l,l,m)$,
appropriate for a given $P$, the general form of a tensor
$x\otimes x^t\otimes z$ such that $x\in L_{j,l}$, $z\in L_{j,m}$,
and $z^t=z$, in the last column of the table. This proves the
theorem for the groups of the form $P\times Z_2$.

It remains to consider the groups $H=Q_i=D_8\circ_i Z_2$,
$i=1,2,3$. We have
$Q_i=(P_i\times1)\leftthreetimes\langle(h_i,(12))\rangle_2$, by
Proposition 3, where $P_{1,2,3}=V^{(1)}$, $V^{(2)},Z_4$, and
$h_{1,2,3}=(13)(24),(12),(12)$, respectively. Each $Q_i$-invariant
decomposable tensor is $P_i\times1$-invariant also, an thus we can
define its type, with respect to the spaces $L_{j,l}$, where $j$
is the number such that $K_j=P_i$, where $K_j$ as in
Proposition~5. By Proposition 10.1, a $Q_i$-invariant decomposable
tensor is of the form $x\otimes R_ix\otimes z$, where $R_i\colon
v\mapsto(h_iv)^t$ and $R_iz=z$.

Note that the transformation $h_i$, and therefore $R_i$, permutes
the spaces $L_{j,l}$: $R_i(L_{j,l})=h_i(L_{j,l})=L_{j,\widehat
l}$, for some transformation $(l\mapsto\widehat l)\in S_4$. It is
not hard to check that this transformation $l\mapsto\widehat l$ is
the same in all three cases, namely it fixes $1$ and $2$ and
interchanges $3$ with $4$.

It is clear that a tensor of type $(l,m,n)$ goes to a tensor of
type $(\widehat m,\widehat l,\widehat n)$ under action of
$g=(h_i,(12))$. So the type $(l,m,n)$ of an $H$-invariant tensor
must satisfy two conditions: $\chi_{j,l}\chi_{j,m}\chi_{j,n}=1$,
and $m=\widehat l$ and $\widehat n=n$. The types, satisfying these
conditions, are the following:  $(1,1,1)$, $(2,2,1)$, $(3,4,2)$,
$(4,3,2)$ when $P=V^{(1)}$ or $V^{(2))}$; and $(1,1,1)$,
$(2,2,1)$, $(3,4,1)$, $(4,3,1)$ when $P=Z_4$.

Finally, we can take into account the invariance under $N_G(H)$,
similarly to the way how it was done earlier. Namely, under action
of $(h_i,1)$ a tensor of type $(l,m,n)$ goes to a tensor of type
$(\widehat l,\widehat m, \widehat n)$. So there is no need to
consider the last of the four types (i.e. $(4,3,2)$ or $(4,3,1)$).

Thus, in all the cases the orbits have representatives of the form
$x\otimes R_ix\otimes z$, where one of the following three
conditions holds: (a) $x,z\in L_{j,1}$, $R_iz=z$; (b) $x\in
L_{j,2}$, $z\in L_{j,1}$, $R_iz=z$, or (c) $x\in L_{j,3}$, $z\in
L_{j,2}$ when $P_i=V^{(1)},V^{(2)}$, $z\in L_{j,1}$ when
$P_i=Z_4$, and $R_iz=z$. To check that in all the cases the
explicit form of the tensor $x\otimes R_ix\otimes z$ coincides
with the corresponding table entry is left to the reader.
\medskip

{\bf 10. The other orbits.} In this section we treat the remaining
cases, namely $S_3\times Z_2$, $A_4\times1$, $A_4\times Z_2$,
$A_4\circ Z_3$, $S_4\times 1$, $S_4\times Z_2$, $S_4\circ Z_2$,
and $S_4\circ_2S_3$.

First let $H=S_4\times1$ or $S_4\times Z_2$. The unique (up to a
scalar) semiinvariant for $S_4$ in $M$ is
$\delta=e_{11}+e_{22}+e_{33}$. So the unique invariant of
$S_4\times1$ or $S_4\times Z_2$ in $M^{\otimes3}$ is
$\delta^{\otimes3}$. But this tensor is invariant under a larger
group $S_4\times S_3$, a contradiction.

Next consider $H=A_4\times1$, $A_4\times Z_2$, and $S_4\circ Z_2$.
All these three groups contain $A_4\times1$ as a normal subgroup.
The semiinvariants of $A_4$ in $M$ are $\delta$,
$\eta=e_{11}+\zeta e_{22}+\overline\zeta e_{33}$, and
$\overline\eta=e_{11}+\overline \zeta e_{22}+\zeta e_{33}$. And
they belong to distinct characters in $A_4^\ast\cong Z_3$. So the
invariants of $A_4\times1$ in $M^{\otimes3}$ are, up to
proportionality, $\alpha_1\otimes\alpha_2\otimes\alpha_3$, where
$\alpha_i\in\{\delta,\eta, \overline\eta\}$ and either
$\alpha_1=\alpha_2=\alpha_3$ or
$\{\alpha_1,\alpha_2,\alpha_3\}=\{\delta,\eta,\overline\eta\}$. In
the case where $\alpha_1=\alpha_2=\alpha_3$ we have
$w=\delta^{\otimes3}$, $\eta^{\otimes3}$, or
${\overline\eta}^{\otimes3}$. These $w$ correspond to groups
$S_4\times S_3$ and $A_4\times S_3$, respectively, so they are not
suitable. On the other hand, all the tensors with
$\{\alpha_1,\alpha_2, \alpha_3\}=\{\delta,\eta,\overline\eta\}$
form an orbit under $1\times S_3$. (Note that $\delta$, $\eta$,
and $\overline\eta$ are symmetric, so the action of $S_3$ is the
permutations of factors without transposing.) As the action of
$(12)\in S_4$ interchanges $\eta,\overline\eta\in M$, the latter
orbit is invariant under $S_4\times1$ and so is a $G$-orbit.

It remains to find out to which of the groups $A_4\times1$,
$A_4\times Z_2$, or $S_4\circ Z_2$ this orbit corresponds. The
group $A_4\times1$ is not appropriate evidently, because its index
$12$ is not equal to 6, the length of the orbit. On the other
hand, the element $(12)\in S_4$ interchanges $\eta$ with
$\overline\eta$, so $w=\eta\otimes\overline\eta\otimes\delta$ is
invariant under $g_2=((12),(12))$. Therefore this $w$ is invariant
under $\langle A_4\times1, g_2\rangle=S_4\circ Z_2$ (and
$A_4\times1$ and $A_4\times Z_2$ are impossible as $H$).

Next consider $H=S_3\times Z_2$. As before, for a decomposable
$S_3\times1$-invariant tensor $w$ we can consider its type
$(l,m,n)$. The condition of $S_3\times1$-invariance immediately
implies that $\{l,m,n\}=\{1,1,1\}$ or $\{2,2,1\}$ (note that
$S_3^\ast\cong Z_2$). It follows from the invariance under
$1\times Z_2$ that $l=m$. So $(l,m,n)=(1,1,1)$ or $(2,2,1)$. Each
$1\times Z_2$-invariant tensor of type $(1,1,1)$ is $x\otimes
x^t\otimes y$, where $x,y\in L_{7,1}$ and $y^t=y$. However, as all
the tensors in $L_{7,1}$ are symmetric, the latter condition can
be rewritten as $w=x\otimes x\otimes y$, $x,y\in L_{7,1}$.
Similarly, an $S_3\times Z_2$-invariant tensor of type $(2,2,1)$
is of the form $\tau\otimes\tau\otimes x$, where $x\in L_{7,1}$
(note that the space $L_{7,2}$ is one-dimensional).

It remains to consider two groups $A_4\circ Z_3$ and
$S_4\circ_2S_3$. By Proposition~3,
$$ A_4\circ Z_3=(V^{(2)}\times1)\leftthreetimes\langle g_5\rangle_3\,,
   \qquad S_4\circ_2 S_3= (V^{(2)}\times1)\leftthreetimes\langle g_2,g_5
   \rangle\,,$$
where $g_2=((12),(12))$ and $g_5=((123),(123))$.

We consider first the larger group $S_4\circ_2S_3$. By
Proposition~11, any $\langle g_2,g_5\rangle$-invariant
decomposable tensor is of the form
$$ (123)x\otimes(132)x\otimes x \eqno (1)$$
where $x$ satisfies condition $Rx=x$, $R\colon v\mapsto((12)v)^t$.
Next, the $V^{(2)}\times1$-invariance of the latter tensor implies
that $x\in L_{5,j}$ for some~$j$. Note that $R$ leaves the spaces
$L_{5,1}$ and $L_{5,2}$ invariant, and interchanges $L_{5,3}$ with
$L_{5,4}$. So $x\in L_{5,1}$ or $L_{5,2}$. But, $R$ acts
identically on $L_{5,2}$, and for such $x$ the tensor (1) is of
the form $w_{44}(a,b)$. Also, the subspace of $R$-invariants in
$L_{5,1}$ is $\langle e_{11}+e_{22}, e_{33}\rangle$, and for $x$
in this subspace the tensor (1) is $w_{43}(a,b)$. This proves the
theorem for $S_4\circ_2S_3$.

Now consider $H=A_4\circ Z_3$. Again, $g_5$-invariant decomposable
tensor must be of the form (1), but not necessarily $Rx=x$. Note
that $(123)\in S_4$ permutes the subspaces $L_{5,j}$ as
$L_{5,1}\mapsto L_{5,1}$, $L_{5,2}\mapsto L_{5,4}\mapsto
L_{5,3}\mapsto L_{5,2}$. So the type of $w$ with respect to
$V^{(2)}\times1$ is one of $(1,1,1)$, $(2,4,3)$, $(3,2,4)$, or
$(4,3,2)$. Note that $1\times Z_3$ normalizes $H$ (even commute
with it elementwise), and cyclically permutes the three latter
types. So we can assume that the type of $w$ is either $(1,1,1)$
or $(4,3,2)$. If the type is $(4,3,2)$, then $x\in L_{5,2}$,
whence $w$ is invariant under $S_4\circ_2 S_3$, a contradiction.
The only type left is $(1,1,1)$. It corresponds to the tensors of
the form $w_{42}(a,b,c)$.

The proof of Theorem 4 is now complete. \smallskip

{\bf Acknowledgement.} The author thanks I.D.Suprunenko for
support and encouragement to publish this work.

(The article was submitted to the journal approximately a year ago. 
Three months ago Irina Dmitrievna had passed away, and I dedicate 
this text to her memory.)

\medskip

\begin{center}\textbf{References}\end{center}
\medskip

1. Aho A.V., Hopcroft J.E., Ullman J.D., The Design
and Analisys of Computer Algorithms. Addison-Wesley, 1974.

2. B\"urgisser P., Clausen M., and Shokrollahi M.A., Algebraic 
Complexity Theory. Springer, 1997.

3. Burichenko V.P., On symmetries of the Strassen algorithm //
arXiv: 1408.6273, 2014. 

4. Burichenko V.P., Symmetries of matrix multiplication algorithms. I 
// arXiv: 1508.01110, 2015.

5. Chiantini L., Ikenmeyer C., Landsberg J.M., Ottaviani G., 
The geometry of rank decompositions of matrix
multiplication I: $2\times2$ matrices // Experimental Mathematics.
28:3 (2019), 322--327. 

6. Burichenko V.P., The isotropy group of the matrix multiplication tensor 
// Trudy Instituta matematiki (= Proceedings of the Institute of
Mathematics). 24:2 (2016), 106--118. See also arXiv: 2210.16565, 2022.

7. Grochow J.A., Moore C., Matrix multiplication algorithms from group orbits 
// arXiv 1612.01527v1. 2016. 

8. Ballard G., Ikenmeyer C., Landsberg J.M., Ryder N., The
geometry of rank decompositions of matrix multiplication II:
$3\times3$ matrices // J.Pure Appl. Algebra, 223:8 (2018), 
3205--3224. 

9. Chokaev B.V., Shumkin G.N., Dva bilineinyh algoritma umnozheniya 
matric $3\times3$ slozhnosti $25$ // Vestn. Mosk.univ. Ser.15. 
Vychisl. matem. i kibern. 2018. Vyp.1. S.23--31 (in Russian). 

9a. Chokaev B.V., Shumkin G.N., Two bilinear $(3\times3)$-matrix 
multiplication algorithms of complexity $25$ // Moscow University 
Computational Mathematics and Cybernetics, 42 (2018), 23--30 
(translation of [9]).  

10. Bl\"aser M., On the complexity of the multiplication of matrices 
of small formats // J.Complexity 19 (2003), 43--60. 

11. Hall,M., The Theory of Groups. Macmillan, 1959.

\end{document}